\documentclass[cits]{PoS}

\newcommand{\be}{\begin{equation}}
\newcommand{\ee}{\end{equation}}
\newcommand{\bea}{\begin{eqnarray}}
\newcommand{\eea}{\end{eqnarray}}

\newcommand{\E}{\mathsf{e}}
\newcommand{\I}{\mathsf{i}}
\newcommand{\srm}[1]{{\textit{\scriptsize #1}}}

\newcommand{\FC}{\;,}

\newcommand{\ot}{\ensuremath{\frac{1}{2}}}

\title{Excited hadrons in $n_f=2$ QCD}

\ShortTitle{Excited hadrons in $n_f=2$ QCD}

\author{G. Engel,$^a$
C. Gattringer,$^a$
\speaker{C. B. Lang},$^a$
M. Limmer,$^a$
D. Mohler,$^a$ and
A. Sch\"afer$^b$
\\
\llap{$^a$}Inst. f. Physik, FB Theoretische Physik, 
Karl-Franzens-Universit\"at Graz, Graz, Austria\\
\llap{$^b$}Inst. f. Theoret. Physik, 
Univ. Regensburg, Regensburg, Gernmay\\
E-mail: 
\email{georg.engel@uni-graz.at},\,\,\email{christof.gattringer@uni-graz.at},
\email{christian.lang@uni-graz.at},\,\,\email{markus.limmer@uni-graz.at},
\email{daniel.mohler@uni-graz.at},\,\,\email{andreas.schaefer@physik.uni-regensburg.de}}

\abstract{
The chirally improved (CI) fermion action allows us to obtain results for pion
masses down to 320 MeV on (in lattice units) comparatively small lattices with
physical extent of 2.4 fm. We use differently smeared quarks sources to build
sets of several interpolators. The variational method then leads to excellent
ground state masses for most mesons and baryons. The excited state signals
weaken in quality towards smaller quark masses. In particular the excited
baryons come out too high. }

\FullConference{The XXVII International Symposium on Lattice Field Theory -
		 LAT2009\\ July 26-31 2009\\ Peking University, Beijing, China}

\begin{document}

\section{Study with two dynamical CI fermions}

We are presenting results of a hadron mass spectrum calculation with emphasis
on possible identification of excited states. The gauge configurations have
been obtained with dynamical,  mass degenerate up and down quarks. For the
fermions we used the chirally improved (CI) Dirac operator $D_{CI}$
\cite{Ga01a,GaHiLa00}. This is a parameterized fermion action of the form
\be
D_{mn}=\sum_{\alpha=1}^{16} \Gamma_\alpha\sum_{p\in{\mathcal P}^\alpha_{m,n}}c_p^\alpha 
\prod_{l\in p}\,U_l\, \delta_{n,m+p}\FC
\ee
where ${\mathcal P}^\alpha_{m,n}$ symbolizes paths from site $m$ to $n$.
Inserting the ansatz in the Ginsparg-Wilson (GW) equation, truncating the
length of the contributions (to, e.g., distance 4), and comparing the
coefficients, leads to a set of algebraic equations, which can be solved (norm
minimization). We used a truncated action with terms involving coupling to
neighboring sites within a $3^4$ hypercube plus some extra terms, giving rise
to several hundred terms. Part of our definition of the  Dirac action is one
step of stout smearing \cite{MoPe04} of the gauge configuration. The
eigenvalues of $D_{CI}$ are closer to the unit circle (where the eigenvalues of
exact GW-operators are located) than those of, e.g., the improved
Wilson operator for the same lattice size (in lattice units). However,
(lattice) chiral symmetry is still violated, albeit to a smaller amount. The
gauge action used is a tadpole-improved  L\"uscher-Weisz action.

The dynamics was implemented with the hybrid Monte Carlo (HMC) algorithm with
Hasenbusch mass preconditioning (with two pseudofermions) and a chronological
inverter utilizing the mixed precision technique \cite{DuFoHo08}. Details and
parameters of the action and the methods of simulation can be found in
\cite{GaHaLa08}. There one also finds a discussion of equilibration and
determination of lattice spacing and AWI (PCAC) mass  of the quarks.

The analysis presented here is based on three ensembles of gauge configurations
for lattice size $16^3\times 32$, with parameters summarized in Table
\ref{tab_summary}.

\begin{table}[htb]
\begin{center}
\begin{tabular}{crrrrrrr}
\hline
\hline
set&	$\beta_\srm{LW}$&$a \,m_0$& $t_\srm{MD}$& configs.& $a$[fm]& $m_\pi$[MeV]& $m_\srm{AWI}$[MeV] \\
\hline
A&	4.70&  -0.050&	 600&100 &0.151(2)&525(7)                         & 42.8(4)\\
B&	4.65&  -0.060&	1200&200 &0.150(1)&470(4)                         & 34.1(2)\\
C&	4.58&  -0.077&  1200&200 &0.144(1)&322(5)                         & 15.3(4)\\
\hline
\end{tabular}
\end{center}
\caption{Overview of the three ensembles of gauge configurations on which this
analysis is based. The parameters given are: gauge coupling, bare mass
parameter $a \,m_0$, number of molecular dynamics (MD) time units, number of
analyzed configurations (in equilibrium and each separated by 5 MD units),
lattice spacing $a$ determined from the static potential with Sommer parameter
0.48 fm, pion mass, AWI mass (from the PCAC relation). For further details see
\cite{GaHaLa08}.}\label{tab_summary}
\end{table}

\section{Variational analysis and hadron interpolators}

In the variational method \cite{Mi85LuWo90} one studies the cross-correlation matrix of several lattice
operators $O_i$ with the correct quantum numbers. 
Inserting a complete set of states (and assuming a discrete spectrum, which
is always the case for finite lattices) one finds a superposition of exponentially
decaying contributions,
\be
C(t)_{ij}=\langle O_i(t)\overline{O_j}(0)\rangle=\sum_n\langle 0|O_i|n\rangle
\langle n| O_j^\dagger | 0\rangle \E^{-t\,M_n}\;.
\ee
Assuming that we have a complete enough set of operators the solution of
the generalized eigenvalue problem
\be
C(t)\,\vec{v}_i=\lambda_i(t)\,C(t_0)\,\vec{v}_i
\ee
allows one to disentangle the individual states and the corresponding energies,
\be
\lambda_i(t)\propto \E^{-t\,M_i}\left(1+\mathcal{O}\left(\E^{-t\,\Delta M_i}\right)\right)\;.
\ee
The eigenvectors are ``fingerprints'' of the states which one may follow through several 
time slices in order to ensure that the state has been identified consistently
(see also the discussion in \cite{BlDeHi09}.)

In order to increase the number of hadron operators and to improve the
correlation signal quality we built interpolating fields with different
smearing of the quark fields. We used Jacobi smeared quark sources, e.g., $u_s\equiv S_s \,u$, with 
an hermitian smearing operator $S_s$ as discussed in \cite{GaHaLa08}, with parameters adjusted to produce two different smearing widths, a wide source ($s=w$, radius 0.55 fm) and a narrow source ($s=n$, radius 0.27 fm).

We also used derivative quark sources as discussed in \cite{GaGlLa08}, e.g., 
sources like $u_{\partial_k}= D_k\,S_w\,u$,  where $k$ denotes the spatial
direction of the covariant derivative
\be
D_i(\vec{x},\vec{y})= \displaystyle U_i(\vec{x},0)
\delta(\vec{x}+\hat{i},\vec{y})-U_i(\vec{x}-\hat{i},0)^\dagger
\delta(\vec{x}-\hat{i},\vec{y})\;.
\ee

The interpolating field operators are built on (in the 3D time slices) 3 times
HYP smeared gauge configurations  \cite{HaKn01}  with smeared valence quark
sources $u_s$, $d_s$ (and the strange quark $s_s$).  We regularly shift the
center of the sources when passing from one configuration to the next in order to improve
decorrelation.

For the meson operators we use bilinears.  Depending on the quantum numbers 
this allows for sets of operators with different Dirac structure and varying
spatial extent. As an example, for the pseudoscalar meson we have
\bea
\overline{u}_n\,\gamma_5\, d_n\;,\;\;
\overline{u}_n\,\gamma_5\, d_w\;,\;\;
\overline{u}_w\,\gamma_5\, d_w\;,\;\;
\overline{u}_n\,\gamma_t\gamma_5\, d_n\;,\;\;
\overline{u}_n\,\gamma_t\gamma_5\, d_w\;,\;\;
\overline{u}_w\,\gamma_t\gamma_5\, d_w\;,\;\;\nonumber\\
\overline{u}_{\partial_i}\,\gamma_i\gamma_5\, d_n\;,\;\;
\overline{u}_{\partial_i}\,\gamma_i\gamma_5\, d_w\;,\;\;
\overline{u}_{\partial_i}\,\gamma_i\gamma_t\gamma_5\, d_n\;,\;\;
\overline{u}_{\partial_i}\,\gamma_i\gamma_t\gamma_5\, d_w\;,\;\;
\overline{u}_{\partial_i}\,\gamma_5\, d_{\partial_i}\;,\;\;
\overline{u}_{\partial_i}\,\gamma_t\gamma_5\, d_{\partial_i}\;.
\eea
The nucleon interpolators have the form
\be
N^{(i)}=\epsilon_{abc} \,\Gamma^{(i)}_1 \,u_a\,\left( u^T_b \,
\Gamma^{(i)}_2\, d_c - d^T_b \,\Gamma^{(i)}_2\, u_c\right)
\ee
with the choices ($\Gamma^{(i)}_1$, $\Gamma^{(i)}_2$) = 
($\mathbf{1}$, $C \gamma_5$),
($\gamma_5$, $C$), and 
($\I\mathbf{1}$, $C \gamma_4\gamma_5$) for $i=1,2,3$ respectively,
where $u$, $d$ denote again 
smeared quarks. The $\Delta$ interpolator is
\be\label{eq:delta_def}
O_{\Delta,k} = \epsilon_{abc}\, u_a\, \big(u_b^T\, C\, \gamma_k\, u_c \big)\ ,
\quad k=1,\,2,\,3\ .
\ee
projected to spin $\frac{3}{2}$. The baryon propagators are projected to
definite parity.

\section{Results for mesons}

\paragraph*{$\mathbf{0^{-+}:\,\pi^\pm(140),\;\pi^\pm(1300)}$.}

As discussed in \cite{GaHaLa08}, in the multi-operator (variational) analysis at
\textit{small} pion masses the backwards running (in time) pion limits the
observation range for the excited state. This can be cured by a larger
time-size; however, for physical pion masses we expect that one needs at least
$N_t=64$ for lattice spacing $a=0.15$ fm or $N_t=128$ for $a=0.075$ fm.

\begin{figure}[t]
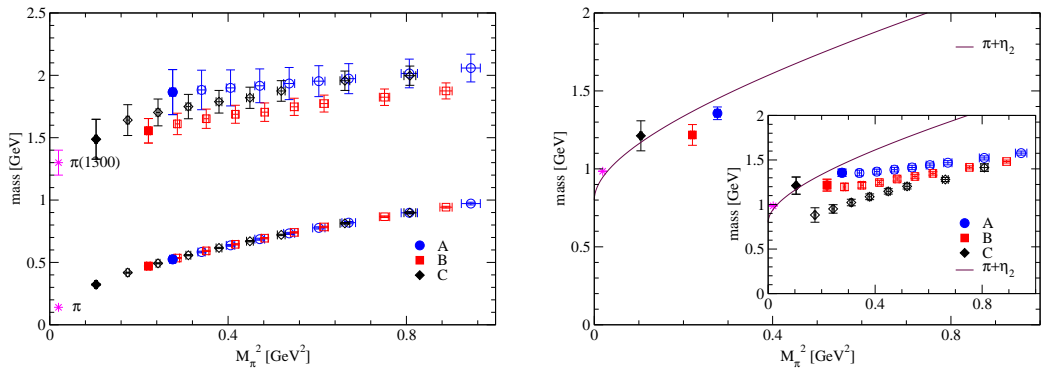

\centering{
\includegraphics[width=6.5cm,clip]{massplot_001000010010.0-+.Ct3-7.eps}
\hfil
\includegraphics[width=6.5cm,clip]{mass_a0_pieta.eps}
}
\caption{\label{fig:pion_and_a0}L.h.s.:  The ground state pion and its first
excitation. Open symbols denote partially quenched data, the full symbols
denote the dynamical data for the three  ensembles A, B and C. R.h.s.: The
dynamical data for the $0^{++}$ channel $a_0$ are compared with the energy
level expected  for a $\pi-\eta_2$ channel. In the insert also the partially
quenched data (open symbols) are shown, as discussed in the main text.}
\end{figure}

We choose the fit interval of the exponential fit to the non-leading eigenvalues
based on the window where the backwards running contribution is not yet dominant
and get the excited pion signal in  Fig. \ref{fig:pion_and_a0} (left). We
include in this plot (as in some of the other figures) the results for partially
quenched data, i.e., where the valence quark masses are larger than the sea
quark masses.

\paragraph*{$\mathbf{0^{++}:\, a_0(980),\;a_0(1450)}$.}

The isovector, scalar meson  has led to controversial results in lattice
simulations \cite{Mc07a,JaMcMi09}.  Most quenched studies found a ground state
extrapolating towards the mass of the $a_0(1450)$ for smaller valence quark
masses. Results for (two) dynamical quarks seem to lead to smaller masses,
compatible with an extrapolation towards the $a_0(980)$. However, for these
masses the energy values are close to those of an expected $\pi-\eta_2$ channel
in s-wave (mass of $\eta_2$ estimated \cite{JaMiUr08}). Fig.
\ref{fig:pion_and_a0} (r.h.s.) exhibits the situation.

When also plotting partially quenched values, we find an interesting effect for
the ensemble C with the smallest sea quark mass. The partially quenched data do
not smoothly extrapolate to the dynamical point. An explanation has been
offered in \cite{PrDaIz04}: the partially quenched states may couple to pairs
of pseudoscalars  (composed of valence and sea quarks), leading  to unphysical
contributions that cancel in the fully dynamical case.

Also, we find a broad range of values for the extracted energy levels for
different sets of contributing interpolators. The issue, whether the $a_0$
ground state is dominantly a tetraquark state, is still not settled (see
\cite{Pr09}).

\paragraph*{$\mathbf{1^{--}:\,\rho^\pm(770),\,\rho^\pm(1450)}$.}

\begin{figure}[t]
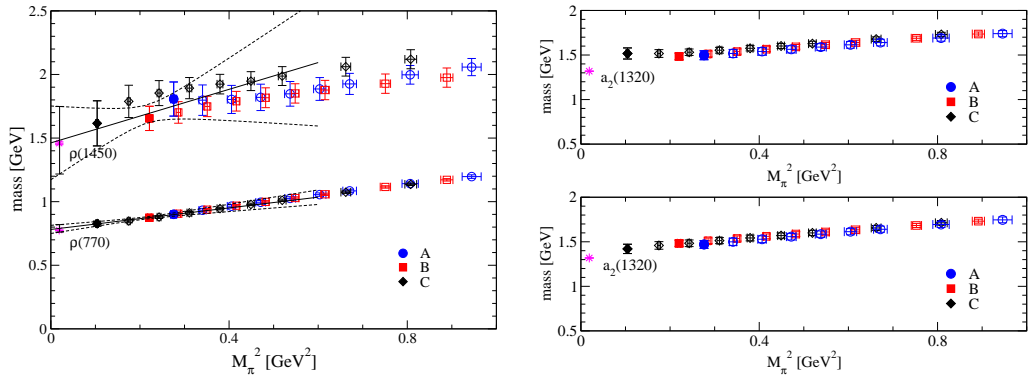

\makebox[\textwidth][t]{ 
\hfil
\vbox{\hsize=6.5cm
\begin{center}
\includegraphics[width=6.5cm,clip]{mass_vector_combined.eps}
\end{center}
}
\hfil
\vbox{\hsize=6.5cm
\begin{center}
\includegraphics[width=6.5cm,clip]{mass_2++_E.eps}\\
\includegraphics[width=6.5cm,clip]{mass_2++_T2.eps}
\end{center}
}
\hfil
}
\caption{
\label{fig:rho} L.h.s.: Ground and excited state in the $1^{--}$ channel. Open
symbols denote partially quenched results.
\label{fig:2++} R.h.s.: Ground state in the $2^{++}$ channel separately
evaluated for interpolator sets in the $E$ (top) and the $T_2$ representation
(below) of the cubic group. Filled symbols denote the values from the three
fully dynamical data sets, open symbols show partially quenched results.}
\end{figure}

In Fig. \ref{fig:rho} (l.h.s.) we show the results for ground state and first excitation
in the $\rho$-channel. For a better signal we allow for different operator
combinations for the excited states. The broken lines give an error window for
linear extrapolating fits based on the three dynamical points only. The
agreement with the experimental values is surprisingly good in view of the
$\rho$ being a resonance; however, the decay pions are in relative p-wave
and due to the given lattice size the necessary extra unit of momentum
stabilizes the vector meson.

We find more admixtures of still higher excitations, and for some combinations
of operators we can identify a 3rd energy level compatible with the
$\rho(1700)$. 

Our data in the exotic $1^{-+}$ channel is too noisy to allow extraction of a
ground state energy level.

\paragraph*{$\mathbf{1^{++}:\, a_1(1260)\textrm{ and }1^{+-}:\,b_1(1235)}$.}

In both channels we see reliable signals only when including derivative
interpolators (cf. the discussion for the quenched case in \cite{GaGlLa08}). The
error bars are somewhat larger for  $1^{+-}$ than for  $1^{++}$, where they are
roughly $\pm 50$ MeV. In both cases a linear extrapolation points towards the
experimental value (within the errors). The mass difference
$m_{latt,b_1}-m_{latt,a_0}$ is approximately 200 MeV.

\paragraph*{$\mathbf{2^{++}:\,a_2(1320)}$.}

In this channel two representations of the cubic group couple: $E$ and $T_2$.
Again interpolators involving derivative sources are necessary, e.g., 
$\epsilon_{ijk}\,\overline{u}_{\partial_i}\,\gamma_j\, d_n$. We find good
signals only when we include several interpolators and thus this is a case
where  the variational method is crucial.  Fig. \ref{fig:2++} (r.h.s.)
demonstrates the situation. The results for the independent analysis of both
representations are compatible, although $T_2$ appears to extrapolate better to
the physical value.

\paragraph*{$\mathbf{2^{-+}:\,\pi_2(1670)\textrm{ and
}2^{--}:\,\rho_2(1940)?}$.}

Qualitatively the situation is like for the $2^{++}$ channel. In both
representations we find values compatible with each other and within large
errors of order 200 MeV also (in linear extrapolation) with the physical mass
value.

\begin{figure}[t]
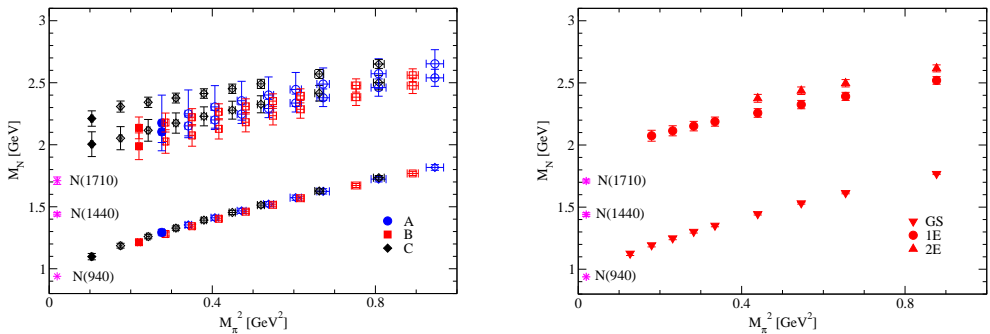

\begin{center}
\includegraphics[width=6cm,clip]{duu_b1_pos_pap.eps}\hfil
\includegraphics[width=6cm,clip]{nuc_M_vs_m_pos_16only.eps}
\end{center}
\caption{\label{fig:nucleon_roper}Results for the positive parity nucleon channel
for the new data with dynamical fermions (l.h.s.) compared to earlier results from 
quenched configurations  \cite{BuGaGl06b} (r.h.s.). 
}
\end{figure}

\section{Results for baryons}
\paragraph*{$\mathbf{\ot^+:\,N(940),\,N(1440),\,N(1710)}$.}

The baryon system poses various challenges in the excited sector, among them the
identification of the Roper state in lattice calculations. In most calculations in
the quenched approximation the first excitation of the (positive parity)
nucleon  came out too high, and only within large error bars an extrapolation
towards the experimental Roper mass was imaginable (see, e.g., the quenched
study in  \cite{BuGaGl06b}). Only recently, and at this conference
\cite{MaCaKa09,Ma09a}, there are results showing an energy level extrapolating
towards smaller masses. That study was based on optimizing combinations of
interpolators with several widths, like in this  and earlier work
\cite{BuGaGl04a}.

There was some hope that introducing the quark dynamics might improve the
situation. Our present analysis does not strengthen that hope (Fig.
\ref{fig:nucleon_roper}). Indeed our new results for dynamical fermions are
similar to earlier quenched results \cite{BuGaGl06b}. We want to emphasize that
it is important in this case to simultaneously find both excitations (the Roper
and the state extrapolating to the $N(1710)$) in order to have a convincing
identification. Both excited states are too high in our results which might
indicate that the given volume is too small for baryon excitations.

\paragraph*{$\mathbf{\ot^-:\,N(1535),\,N(1650)}$.}

We clearly identify two states but cannot quantify their mass splitting due to
large errors.

\paragraph*{$\mathbf{\frac{3}{2}^+:\,}\Delta\mathbf{(1232),\,}\Delta\mathbf{(1600)\textrm{
and }\frac{3}{2}^-:\,}\Delta\mathbf{(1700)}$.}

\begin{figure}[t]
\begin{center}
\includegraphics[width=6cm,clip]{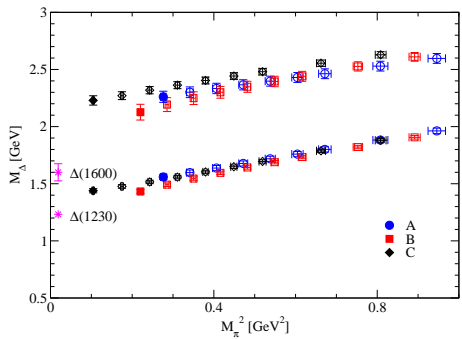} 
\end{center}
\caption{\label{fig:delta_pos}Positive parity: Ground state $\Delta$ and first excitation. 
}
\end{figure}

In the positive parity sector (see Fig. \ref{fig:delta_pos}) we find that the ground state is
closer to the experimental values, as compared to the quenched results. The excited state is
clearly seen, but  again too high, maybe due to volume squeezing. The ground state in the
negative parity sector is clearly identified and extrapolates towards the experimental
mass.

Thus, while our meson results look quite good, the excited baryons come out too high. 
Improvement of the signal might be obtained with larger lattices and possibly extending the set of interpolator (see, e.g., \cite{PeBuFo09}). 
Further results for the axial charge of baryons including baryons with strangeness are discussed in \cite{Mo09}.

{\bf Acknowledgment:} This work has been supported by Austrian FWF  DK W1203-N08 and German DFG project SFB/TR-55. The calculations have been performed on the SGI Altix 4700 of the Leibniz-Rechenzentrum Munich and on local clusters at ZID at the University of Graz.
    
\providecommand{\href}[2]{#2}\begingroup\raggedright\endgroup


\begin{thebibliography}{10}

\bibitem{Ga01a}
C.~Gattringer, 
{\em
  Phys. Rev. D} {\bf 63} (2001) 114501
  [\href{http://arXiv.org/abs/hep-lat/0003005}{{\tt hep-lat/0003005}}].

\bibitem{GaHiLa00}
C.~Gattringer, I.~Hip and C.~B. Lang,  
  {\em Nucl. Phys. B} {\bf 597} (2001) 451
  [\href{http://arXiv.org/abs/hep-lat/0007042}{{\tt hep-lat/0007042}}].

\bibitem{MoPe04}
C.~Morningstar and M.~Peardon,  
  {\em Phys. Rev. D} {\bf 69} (2004) 054501
  [\href{http://arXiv.org/abs/hep-lat/0311018}{{\tt hep-lat/0311018}}].

\bibitem{DuFoHo08}
S.~D\"urr, Z.~Fodor, C.~Hoelbling, R.~Hoffmann, S.~Katz, S.~Krieg, T.~Kurth,
  L.~Lellouch, T.~Lippert, K.~Szabo and G.~Vulvert,  
   {\em Phys. Rev. D} {\bf 79} (2008)
  014501 [\href{http://arXiv.org/abs/arXiv:0802.2706 [hep-lat]}{{\tt
  arXiv:0802.2706 [hep-lat]}}].

\bibitem{GaHaLa08}
C.~Gattringer, C.~Hagen, C.~B. Lang, M.~Limmer, D.~Mohler and A.~Sch\"afer,
  {\em Phys. Rev. D} {\bf 79} (2009) 054501
  [\href{http://arXiv.org/abs/0812.1681}{{\tt 0812.1681}}].

\bibitem{Mi85LuWo90}
C.~Michael,  
{\em  Nucl. Phys. B} {\bf 259} (1985) 58.
M.~L{\"u}scher and U.~Wolff,  
  {\em Nucl. Phys. B} {\bf 339} (1990) 222.

\bibitem{BlDeHi09}
B.~Blossier, M.~DellaMorte, G.~von Hippel, T.~Mendes and R.~Sommer,  
  {\em JHEP} {\bf 0904} (2009) 094
  [\href{http://arXiv.org/abs/0902.1265}{{\tt 0902.1265}}].

\bibitem{GaGlLa08}
C.~Gattringer, L.~Y. Glozman, C.~B. Lang, D.~Mohler and S.~Prelovsek,
  {\em Phys.
  Rev. D} {\bf 78} (2008) 034501 
  [\href{http://arXiv.org/abs/arXiv:0802.2020
  [hep-lat]}{{\tt arXiv:0802.2020 [hep-lat]}}].

\bibitem{HaKn01}
A.~Hasenfratz and F.~Knechtli,  
  {\em Phys. Rev. D} {\bf 64}
  (2001) 034504
  [{\tt arXiv:hep-lat/0103029}].
 
\bibitem{Mc07a}
C.~McNeile,  
{\em PoS} ({\bf LATTICE2007})
  (2007) 019 [\href{http://arXiv.org/abs/arXiv:0710.0985 [hep-lat]}{{\tt
  arXiv:0710.0985 [hep-lat]}}].

\bibitem{JaMcMi09}
K.~Jansen, C.~McNeile, C.~Michael and C.~Urbach, (2009)
[\href{http://arXiv.org/abs/arXiv:0906.4720
  [hep-lat]}{{\tt arXiv:0906.4720 [hep-lat]}}].

\bibitem{JaMiUr08}
K.~Jansen, C.~Michael and C.~Urbach, 
  {\em Eur.\ Phys.\ J.\  C} {\bf 58} (2008) 261
[\href{http://arXiv.org/abs/arXiv:0804.3871
  [hep-lat]}{{\tt arXiv:0804.3871 [hep-lat]}}].

\bibitem{PrDaIz04}
S.~Prelovsek, C.~Dawson, T.~Izubuchi, K.~Orginos and A.~Soni, 
  {\em Phys. Rev. D} {\bf 70} (2004) 094503
  [\href{http://arXiv.org/abs/hep-lat/0407037}{{\tt hep-lat/0407037}}].

\bibitem{Pr09}
S.~Prelovsek {\em et.~al.},
contribution to this conference, {\em  PoS} ({\bf LAT2009}) (2009) 088.

\bibitem{BuGaGl06b}
T.~Burch, C.~Gattringer, L.~Y. Glozman, C.~Hagen, D.~Hierl, C.~B. Lang and
  A.~Sch{\"a}fer,  
{\em  Phys. Rev. D} {\bf 74} (2006) 014504
  [\href{http://arXiv.org/abs/hep-lat/0604019}{{\tt hep-lat/0604019}}].

\bibitem{MaCaKa09}
M.~S. Mahbub, A.~O. Cais, W.~Kamleh, B.~G. Lasscock, D.~B. Leinweber and A.~G.
  Williams,  
  {\em Phys. Rev. D} {\bf 80} (2009) 054507
  [\href{http://arXiv.org/abs/arXiv:0905.3616v1 [hep-lat]}{{\tt
  arXiv:0905.3616v1 [hep-lat]}}].

\bibitem{Ma09a}
S.~Mahbub {\em et.~al.}, 
contribution to this conference, {\em  PoS} ({\bf LAT2009}) (2009).

\bibitem{BuGaGl04a}
T.~Burch, C.~Gattringer, L.~Y. Glozman, R.~Kleindl, C.~B. Lang and
  A.~Sch{\"a}fer,  
  {\em Phys. Rev. D} {\bf 70} (2004) 054502
  [\href{http://arXiv.org/abs/hep-lat/0405006}{{\tt hep-lat/0405006}}].

\bibitem{PeBuFo09}
M.~Peardon, J.~Bulava, J.~Foley, C.~Morningstar, J.~Dudek, R.~G. Edwards,
  B.~Joo, H.-W. Lin, D.~G. Richards and K.~J. Juge, 
 (2009)
  [\href{http://arXiv.org/abs/arXiv:0905.2160 [hep-lat]}{{\tt
  arXiv:0905.2160 [hep-lat]}}].

\bibitem{Mo09}
D.~Mohler {\em et.~al.}, contribution to this conference, {\em  PoS} 
({\bf LAT2009}) (2009).

\end{thebibliography}
\end{document}